\documentclass[useAMS,usenatbib]{mn2e}

\usepackage{graphicx}
\usepackage{psfig}
\usepackage{makeidx,amsmath,mathrsfs}

\onecolumn

\newcommand\apj{ApJ}
\newcommand\apjl{ApJ}
\newcommand\apjs{ApJS}
\newcommand\aap{A$\&$A}
\newcommand\mnras{MNRAS}
\newcommand\prd{Phys.~Rev.~D}
\newcommand\sovast{Soviet~Ast.}
\newcommand\nat{Nature}

\title[Initial shear field]{On the initial shear field of the cosmic web}

\author[G. Rossi]
{Graziano Rossi\thanks{Email: graziano@kias.re.kr} \\
\\
\footnotesize Korea Institute for Advanced Study, Hoegiro 87, Dongdaemun-Gu, Seoul $130-722$, Korea}
\date{Accepted 2011 December 01. Received 2011 November 30; in
  original form 2011 July 29}
\pagerange{\pageref{firstpage}--\pageref{lastpage}}
\pubyear{2011}

\begin{document}
\maketitle
\label{firstpage}



\begin{abstract}

The initial shear field, characterized by a primordial perturbation potential,
plays a crucial role in the formation of
large scale structures. Hence, considerable analytic work has been based 
on the joint distribution of its eigenvalues, associated with Gaussian statistics. In addition, directly related
morphological quantities such as ellipticity or prolateness
are essential tools in understanding the formation and structural properties
of halos, voids, sheets and filaments, their relation with the local environment, and the
geometrical and dynamical classification of the cosmic web.
To date, most analytic work has been focused on    
Doroshkevich's unconditional formulae for the eigenvalues of the linear tidal field, which neglect 
the fact that halos (voids) may correspond to
maxima (minima) of the density field. 
I present here new formulae for the constrained eigenvalues of the initial shear field
associated with Gaussian statistics, which include the fact that those eigenvalues are related to
regions where the source of the displacement is positive (negative):
this is achieved by requiring the Hessian matrix of the displacement field to be positive (negative) definite.    
The new conditional formulae naturally reduce to Doroshkevich's unconditional
relations, in the limit of no correlation between the potential and
the density fields. 
As a direct application, I derive the 
individual conditional distributions of eigenvalues and point out the
connection with previous literature. Finally, I outline other possible theoretically- or
observationally-oriented uses, 
ranging from studies of halo and void triaxial formation, development of structure-finding
algorithms for the morphology and topology of the cosmic web, till 
an accurate mapping of the gravitational potential environment of galaxies 
from current and future generation galaxy redshift surveys. 

\end{abstract}



\begin{keywords}
methods: analytical, statistical --- cosmology: theory, large-scale
structure of Universe --- galaxies: formation 
\end{keywords}



\section{Introduction}


The large-scale spatial distribution of dark
matter, as revealed from numerical simulations, shows a
characteristic anisotropic web-like structure. This \textit{cosmic web}, which arises through the gravitational clustering of
matter, is mainly due to the effects of the tidal field: in fact,
the competition between
cosmic expansion, the trace, and the traceless part
of the tidal field imprints anisotropies in the large-scale matter distribution in much
the same way that gravity and radiation pressure imprints baryonic
acoustic oscillations (BAO) on the
Cosmic Microwave Background (CMB) sky (Hu \& Sugiyama 1995; Lee \& Springel
2010).
Hence, the initial shear field plays a crucial role in the formation of large
scale structures, and a number of studies in the literature have been
devoted to this subject -- among the plethora of papers, see for example the
classic works by Zeldovich (1970), Icke (1973), Peebles (1980), White
(1984), Bardeen et
al. (1986), Kaiser (1986), Bertschinger (1987), Bond \& Myers
(1996), Bond, Kofman \& Pogosyan (1996) and van de Weygaert \&
Bertschinger (1996). 
In addition, if the cosmic web
originates from primordial tidal effects and its degree of
anisotropy increases with the evolution of the Universe, 
then studying this initial field is crucial in
understanding the subsequent nonlinear evolution of cosmic structures
(Springel et al. 2005; Shandarin et al. 2006; Desjacques 2008;
Desjacques \& Smith 2008; Pogosyan et al. 2009), the alignment of shape and angular
momentum of halos (West 1989; Catelan et al. 2001; Lee \& Springel 2010; Rossi, Sheth \& Tormen 2011), 
the statistical properties of voids (Lee \&
Park 2006; Platen, van de Weygaert \& Jones 2008),
and more generally for characterizing the geometry and morphology of the cosmic web
(Shen et al. 2006;  van de Weygaert \& Bond 2008; Forero-Romero et al. 2009; Aragon-Calvo et al. 2010a,b; Shandarin et al. 2010). 

The basic theory for the formation and evolution of structures is now
well-understood, thanks to the
pioneering work of 
Doroshkevich \&
Zeldovich (1964),  Doroshkevich (1970), Zeldovich (1970), and Sunyaev \& Zeldovich (1972) --
the latter in the context of galaxy formation.
In particular, Doroshkevich (1970) derived 
the joint probability distribution of an ordered set of eigenvalues in the tidal
field matrix -- at random positions -- given the variance of the density field,
corresponding to a Gaussian potential; we will refer
to it as the \textit{unconditional} probability distribution of eigenvalues. 
In addition, 
Zeldovich (1970) provided the fundamental understanding of anisotropic collapse
on cosmological scales, and recognized the key role of the large scale
tidal force in shaping the cosmic web. Subsequently,
Doroshkevich and Shandarin (1978) calculated some
statistical properties of the maxima of the largest eigenvalue of the
shear tensor. Their study proved that 
the most probable formation process starts first with
a one-dimensional collapse (cosmic pancake formation).
The directions (orientations) for the one-dimensional collapsed sheets
are determined by the largest eigenvalue of the deformation tensor,
which can be attributed to the initial linear density perturbations;
moreover, the probability that two or even three of the initial eigenvalues are identical or nearly equal
is extremely small, indicating that the collapse is triaxial. 
Later on, following a constrained field approach pioneered by 
Bertschinger (1987), van de Weygaert \&
Bertschinger (1996) developed an algorithm for setting up tailor-made
initial conditions for cosmological simulations, which addressed the
role of the tidal fields in shaping large-scale structures.
In the same period, Bond, Kofman \& Pogosyan (1996) developed a cosmic web
theory which naturally explains the filamentary structure present in the Cold
Dark Matter (CDM) cosmology, due to the coherent nature of the
primordial tidal field. They realized that an ``embryonic'' cosmic
web is already present in the primordial density field, and
explained why in overdense regions sheet-like membranes are only
marginal features. 
Since then, because of the correspondence between structures in the evolved density
field and local properties of the linear tidal field pointed out by 
the same authors, the statistics of the shear has
received more attention in the literature. 
For example, Lee \& Shandarin (1998) computed some
probability distributions for individual shear eigenvalues and
obtained an analytic approximation to the halo mass function, and 
Catelan \& Porciani (2001) explored the two-point correlation of
the tidal shear components. 

However, a variety of studies in the physics of halo
formation and cosmic web classification are based on 
Doroshkevich's unconditional formulae for the ordered eigenvalues of the initial
shear field associated with Gaussian statistics (Doroshkevich 1970), but
those formulas cannot differentiate between random positions and peak/dips
as they neglect the fact that 
halos (voids) may correspond to
maxima (minima) of the density field. According to Bardeen et
al. (1986),
if one assumes the cosmological density fluctuations to be Gaussian
random fields, the local maxima of such fields are plausible sites for the formation of
nonlinear structures. Hence, 
the statistical properties of the peaks can be
used to predict the abundances and clustering properties of objects of
various types, and in studies of the non-spherical formation of
large-scale structures. 
The study of Bertschinger (1987) goes in this direction, by
generalizing the treatment of Bardeen et
al. (1986) and proposing
a path integral method for sampling constrained Gaussian random
fields. The method allows one to study the density field around peaks
or other constrained regions in the biased galaxy formation scenario,
but unfortunately it is too elaborate and inefficient in its
implementation. 
Instead, by applying the prescription of Hoffman \& Ribak
(1991) to construct constrained random fields, van de Weygaert \&
Bertschinger (1996) were able to show that it is possible to generate
efficiently initial Gaussian random density and velocity
fields, and specify the presence and characteristics of one or more
peaks and dips at arbitrary locations -- with the gravity and tidal
fields at the site of the peaks having the required strength and orientation. 
Some other studies on constrained initial conditions were
also pursued by van Haarlem \& van de Weygaert (1993), and by van de
Weygaert \& Babul (1994).  
The idea has been expanded in Bond \& Myers (1996),
who presented a \textit{peak-patch} picture of structure formation as
an accurate model of the dynamics of peaks in the density field. 
Their approach goes further, as it involves the explicit formalism for
identifying objects in a multiscale field (i.e. it does not restrict
to a single scale).
The model is even more precise for void patches, the
equivalent framework for studying voids (Sahni et al. 1994; Sheth \& van
de Weygaert 2004; Novikov, Colombi \& Dore 2006; Colberg et al. 2008).
In general, density peaks define a well-behaved point-process which
can account for the discrete nature of dark matter halos and galaxies,
and on asymptotically large scales are linearly biased tracers of the
dark matter field (Desjacques \& Sheth 2010).

Therefore, it would be desirable to incorporate the 
peak (or dip) constraint in the statistical description of the 
initial shear field, in order to characterize more realistically the geometry and dynamics of the cosmic web.
The main goal of this paper 
is to do so, by providing a set of analytic expressions which extend
the work of Doroshkevich (1970) and Bardeen et al. (1986), and are akin in philosophy to that of
van de Weygaert \& Bertschinger (1996).
This is achieved by constraining the Hessian 
of the displacement field (the matrix of the second derivatives) to be positive (negative) definite, which is
the case in the vicinity of minima (maxima) of the source
of the displacement field. 
The new \textit{conditional} probability distributions derived in this
study include the correlation between the potential and
density fields through
a reduced parameter $r$, and naturally recover Doroshkevich's (1970)
\textit{unconditional} formulae
in the absence of correlation. 

Hence, the main focus of this work is to derive explicitly the joint probability distribution of the eigenvalues
of the shear field, given the fact that positions are peaks
or dips in the corresponding density field -- and not random
locations. In this sense 
the field is termed \textit{constrained} (i.e. it has \textit{constrained} eigenvalues,
which are the result of looking only at peak/dip regions), 
and it is statistically described by a
\textit{conditional} probability distribution.
Note that even though the formalism is essentially restricted to one scale, the
formulae derived here are useful in a variety of applications -- some of which will be
discussed at the end of this paper.


The layout is organized as follows. 
Section \ref{new_formula} provides the derivation of the new analytic expressions
for the conditional distributions of eigenvalues of the initial shear field.
In particular, Section \ref{notation} illustrates the basic notation
adopted; Section \ref{joint_eigen_new} contains the derivation of
the joint distribution of eigenvalues in the peak/dip picture, while 
Section \ref{bbks_connection} shows the reverse conditional
distribution function.
As a direct application of the new formulae, 
Section \ref{individual_eigenvalues} presents the  
individual distributions of eigenvalues subjected to the
extremum constraint, along with some other related conditional
probabilities. This part extends previous work by Lee \& Shandarin (1998), which is briefly 
summarized in Appendix \ref{Lee_Shandarin}, and complements the study of
van de Weygaert \& Bertschinger (1996).
Finally, Section \ref{conclusion} highlights the main results and discusses ongoing and future
applications, which will be presented in forthcoming publications.



\section{Joint distribution of eigenvalues at peak/dip locations} \label{new_formula}

In this section, I first introduce the basic notation adopted throughout the paper.
I then  derive the conditional joint distribution of eigenvalues of
the tidal field, under the constraint that the Hessian matrix 
of the source of the displacement is positive (or negative) definite.   
Finally, I present the reverse probability function, useful in
relating this work with that of Bardeen et al. (1986) and van de Weygaert \& Bertschinger (1996).


\subsection{Basic notation} \label{notation}


Let $\Psi$ denote the displacement field, $\Phi$ the potential of the displacement field, $S_{\rm \Psi}$ the source of the displacement field.
Indicate with ${\bf q}$ the Lagrangian coordinate, with ${\bf x}$ the Eulerian coordinate, where
\begin{equation} 
{\bf x} ({\bf q}) = {\bf q} + \Psi({\bf q}).
\end{equation}
Use $T_{\rm ij}$ to denote the shear tensor of the displacement field,
$H_{\rm ij}$ for the Hessian matrix, $J_{\rm ij}$ for the Jacobian of the
displacement field ($i,j=1,2,3$). Clearly:
\begin{equation}
J_{\rm ij}({\bf q}) = { \partial x_{\rm i}  \over \partial q_{\rm j}}
= \delta_{\rm ij} + T_{\rm ij} 
\end{equation}

\begin{equation}
T_{\rm ij} = {\partial \Psi_{\rm i} \over \partial q_{\rm j}} =
{\partial^2 \Phi  \over \partial q_{\rm i} \partial q_{\rm j}}
\end{equation}

\begin{equation}
H_{\rm ij} = {\partial^2 S_{\Psi} \over \partial q_{\rm i} \partial
  q_{\rm j}}
\end{equation}

\begin{equation}
S_{\Psi} ({\bf q}) = \sum_{\rm i=1}^3 {\partial \Psi_{\rm i} \over
  \partial q_{\rm i}} \equiv  \sum_{\rm i=1}^3 {\partial^2 \Phi \over
  \partial q^2_{\rm i}}.
\end{equation} 
The eigenvalues of $T_{\rm ij}$ are $\lambda_1, \lambda_2, \lambda_3$,
those of $H_{\rm ij}$ are $\xi_1, \xi_2, \xi_3$. 
In general, the ordering of the eigenvalues is assumed to be such that
$\lambda_1 \ge \lambda_2 \ge \lambda_3$.
Let the potential $\Phi$ be a Gaussian random field determined by the
power spectrum of matter density fluctuations $P(k)$, with $k$
denoting the wave number and $W(k)$ the smoothing kernel.
The density field described by the source $S_{\Psi}$ is also a Gaussian random field. The correlations between these two fields are expressed by:
\begin{equation}
\langle   T_{\rm ij} T_{\rm kl} \rangle = { \sigma_{\rm T}^2 \over 15}
( \delta_{\rm ij} \delta_{\rm kl}  + \delta_{\rm ik} \delta_{\rm jl} +
\delta_{\rm il} \delta_{\rm jk})  
\end{equation}
 
\begin{equation}
\langle   H_{\rm ij} H_{\rm kl} \rangle = { \sigma_{\rm H}^2 \over 15}
( \delta_{\rm ij} \delta_{\rm kl}  + \delta_{\rm ik} \delta_{\rm jl} +
\delta_{\rm il} \delta_{\rm jk})  
\end{equation}
 
\begin{equation}
\langle   T_{\rm ij} H_{\rm kl} \rangle = { \Gamma_{\rm TH}^2 \over
  15} ( \delta_{\rm ij} \delta_{\rm kl}  + \delta_{\rm ik} \delta_{\rm
  jl} + \delta_{\rm il} \delta_{\rm jk})  
\end{equation}
where $\sigma_{\rm T}^2 = S_2 \equiv \sigma_0^2$, $\sigma_{\rm H}^2 = S_6
\equiv \sigma_2^2$, $\Gamma_{\rm TH} = -S_4 \equiv - \sigma_1^2$,
$\delta_{\rm ij}$ is the Kronecker delta, and
\begin{equation}
S_{\rm n} = {1 \over 2 \pi^2} \int_0^{\infty} k^{\rm n}~P(k)~W^2(k) {\rm d}k 
\end{equation}
 
\begin{equation}
\sigma_{\rm j}^2 = {1 \over 2 \pi^2} \int_{0}^{\infty} k^{\rm
  2(j+1)}~P(k)~W^2(k)~{\rm d}k \equiv S_{\rm 2(j+1)}.  
\end{equation}
$T_{\rm ij}$ and $H_{\rm ij}$ are real symmetric tensors, so they are specified
by 6 components (note however that only 5 are independent once the
height of the density peak or dip has been set, because the trace of
the shear field tensor is entirely constrained by the local density
field value via the Poisson equation). 
Label them as $A=(1,1)$, $B=(2,2)$, $C=(3,3)$, $D=(1,2)$, $E=(1,3)$, $F=(2,3)$;  
the symbols $\alpha$ or $\beta$ indicate the various couples, in a compact notation, where $\alpha, \beta = A,B,C,D,E,F$.
For example, if $\alpha = D$ then $T_{\alpha} = T_{\rm D} \equiv T_{12}$, and so forth.
In the subsequent derivations, for clarity all the various dependencies on $\sigma$'s
are dropped. This is done by introducing the ``reduced''
variables $\tilde{T}$ and $\tilde{H}$, defined as
\begin{equation}
\tilde{T}_{\alpha} = T_{\alpha}/\sigma_{\rm T}, ~~\tilde{\rm H}_{\alpha} = H_{\alpha}/\sigma_{\rm H}
\label{tilde_var}
\end{equation} and the ``reduced'' correlation
\begin{equation}
r = \Gamma_{\rm TH}/\sigma_{\rm T} \sigma_{\rm H} = -{\sigma_1^2 \over \sigma_0 \sigma_2} \equiv = - \gamma
\label{tilde_corr}
\end{equation}
where $\gamma$ is the same as in Eq. (4.6a) of Bardeen et al. (1986).
In this notation, the eigenvalues of $\tilde{T}$ and of $\tilde{H}$
are $\tilde{\lambda}_i$'s and $\tilde{\xi}_i$'s, respectively, with $i=1,2,3$.
Of course, the $\sigma$ dependence can be restored at any time, if desired,
by using the previous Equations (\ref{tilde_var}) and (\ref{tilde_corr}).


\subsection{Joint conditional distribution of eigenvalues in the peak/dip picture} \label{joint_eigen_new}


The main goal of this section is to derive the joint distribution of the eigenvalues of
$J_{\rm ij}$ when the ordering is $\lambda_1 \ge \lambda_2 \ge \lambda_3$, in regions
where the source of displacement $S_{\Psi}$ is a minimum (or maximum).
Being a minimum (maximum) of $S_{\Psi}$ implies that the displacement field is such that the gradient of $S_{\Psi}$ is zero, and the Hessian matrix $\tilde{H}_{\alpha}$ 
is positive (or negative) definite. In block notation, the covariance matrix $\tilde{V}$
of the 12 reduced components of $Y=(\tilde{T}, \tilde{H})$ is:
\begin{equation}
 \tilde{V} = \begin{pmatrix}
  \langle \tilde{T}_{\alpha} \tilde{T}_{\alpha} \rangle &   \langle \tilde{T}_{\alpha} \tilde{H}_{\beta} \rangle \\
    \langle \tilde{H}_{\beta} \tilde{T}_{\alpha} \rangle &   \langle \tilde{H}_{\beta} \tilde{H}_{\beta} \rangle 
 \end{pmatrix} =
 {1 \over 15} \begin{pmatrix}
 A & r A \\
  r A & A
\end{pmatrix}
\end{equation}
where
\begin{equation}
 A = \begin{pmatrix}
  B & \oslash \\
  \oslash &I  
 \end{pmatrix},~
  B = \begin{pmatrix}
  3 & 1 & 1\\
  1 & 3&1\\
  1&1&3 
 \end{pmatrix}
 \end{equation}
with $I$ a ($3\times3$) identity matrix and $\oslash$ a ($3 \times 3$) null matrix.
The inverse of the reduced covariance matrix is simply:
\begin{equation}
 \tilde{C} = \tilde{V}^{-1} =  {15 \over (1 -r^2)} \begin{pmatrix}
  A^{-1} & -r A^{-1} \\
  -r A^{-1} & A^{-1}  
 \end{pmatrix}
 \end{equation}
 where
\begin{equation}
 A^{-1} = \begin{pmatrix}
  B^{-1} & \oslash \\
  \oslash &I  
 \end{pmatrix},~
  B^{-1} = {1 \over 10} \begin{pmatrix}
  4 & -1 & -1\\
  -1 & 4&-1\\
  -1&-1&4 
\end{pmatrix}.
\end{equation} 
The joint probability of observing a tidal field $\tilde{T}$ for the gravitational potential and a curvature $\tilde{H}$
for the density field (from now on, the understood indices $\alpha$
and $\beta$ are dropped) is a multivariate Gaussian in $Y = (\tilde{T}, \tilde{H})$:
\begin{equation}
p(Y|r) {\rm d} Y =  { e^{ - {1 \over 2} Y^T \cdot \tilde{C} \cdot Y}  \over (2 \pi)^6   \sqrt{|\tilde{V}| }}  {\rm d}Y
\equiv p(\tilde{T}, \tilde{H}|r) {\rm d}\tilde{T} {\rm d} \tilde{H}.  
\end{equation}
Therefore
\begin{equation}
p(\tilde{T}| r, \tilde{H} > 0 ) = {\int_{\rm \tilde{H} >0}
 p(\tilde{T}, \tilde{H}|r)~{\rm d}
  \tilde{H} \over p(\tilde{H}>0)}=
  {\int_{\rm \tilde{H}>0} p(\tilde{H}) \cdot
  p(\tilde{T}|r, \tilde{H})~{\rm d} \tilde{H} \over \int_{\rm
 \tilde{H}>0} p(\tilde{H})~{\rm d}\tilde{H}},
\label{what_we_want}
\end{equation}
where
\begin{equation}
p(\tilde{T}|r, \tilde{H}) = { p(\tilde{T}, \tilde{H}|r) \over
  p(\tilde{H})} = { p(\tilde{T}, \tilde{H}|r) \over  \int_{\rm
  \tilde{T}} p(\tilde{T}, \tilde{H}|r)~{\rm d} \tilde{T}} = 
{ e^{ - {1 \over 2} \rm ( \tilde{T}-b)^T \cdot {\mathscr{A}}^{-1} \cdot (\tilde{T}-b) }  \over (2 \pi)^3   \sqrt{|\mathscr{A}| } }.
\label{cond}
\end{equation}
The marginal distributions $p(\tilde{T})$ and $p(\tilde{H})$ are
simply multidimensional Gaussians with covariance matrix $A/15$, and
therefore they can be expressed using Doroshkevich's formulae as
\begin{equation}
p(\tilde{T}) = {15^3 \over 16 \sqrt{5} \pi^3} e^{- \rm {3 \over 2} (2 k_1^2 - 5 k_2)}
\label{doro_standard}
\end{equation}
with
\begin{eqnarray}
k_1 &=& \tilde{T}_{11} + \tilde{T}_{22} + \tilde{T}_{33} \\
k_2 &=&  \tilde{T}_{11}\tilde{T}_{22}+ \tilde{T}_{11}\tilde{T}_{33}+ \tilde{T}_{22}\tilde{T}_{33} -\tilde{T}^2_{12} -\tilde{T}^2_{13} -\tilde{T}^2_{23} 
\end{eqnarray}
and similarly
\begin{equation}
p(\tilde{H}) = {15^3 \over 16 \sqrt{5} \pi^3} e^{- \rm {3 \over 2} (2 h_1^2 - 5 h_2)}
\end{equation}
with
\begin{eqnarray}
h_1 &=& \tilde{H}_{11} + \tilde{H}_{22} + \tilde{H}_{33} \\
h_2 &=&  \tilde{H}_{11}\tilde{H}_{22}+ \tilde{H}_{11}\tilde{H}_{33}+ \tilde{H}_{22}\tilde{H}_{33} -\tilde{H}^2_{12} -\tilde{H}^2_{13} -\tilde{H}^2_{23}. 
\end{eqnarray}
The conditional distribution
$p(\tilde{T}|r, \tilde{H})$ is also a multidimensional Gaussian, with mean $b$ and covariance matrix $\mathscr{A}$, where
\begin{equation}
b = r \tilde{H}, ~ {\mathscr{A}} = {1 \over 15} (1 -r^2) A.
\end{equation}
Carrying on the calculation indicated in (\ref{cond}) yields:
\begin{equation}
p (\tilde{T}|r, \tilde{H}) = {15^3 \over 16 \sqrt {5} \pi^3} {1 \over (1-r^2)^3} {\rm exp} \Big [-{3 \over 2 (1-r^2)} (2 K_1^2 - 5 K_2 ) \Big ]
\label{doro_inter_extended}
\end{equation}
where
\begin{eqnarray}
\label{K_def}
K_1 &=& (\tilde{T}_{11} -r \tilde{H}_{11}) + (\tilde{T}_{22} -r \tilde{H}_{22}) + (\tilde{T}_{33} -r \tilde{H}_{33}) = k_1 - r h_1 \nonumber \\
K_2 &=& (\tilde{T}_{11} -r \tilde{H}_{11})(\tilde{T}_{22} - r \tilde{H}_{22})  + (\tilde{T}_{11} -r \tilde{H}_{11})(\tilde{T}_{33} - r \tilde{H}_{33}) \nonumber \\
&+& (\tilde{T}_{22} -r \tilde{H}_{22})(\tilde{T}_{33} - r \tilde{H}_{33}) -(\tilde{T}_{12} -r \tilde{H}_{12})^2 
-(\tilde{T}_{13} -r \tilde{H}_{13})^2 -(\tilde{T}_{23} -r \tilde{H}_{23})^2 = k_2 + r^2 h_2 - r h_1 k_1 + r \eta
\end{eqnarray}
and
\begin{equation}
\eta = \tilde{T}_{11} \tilde{H}_{11} +\tilde{T}_{22} \tilde{H}_{22} +\tilde{T}_{33} \tilde{H}_{33} + 2 \tilde{T}_{12} \tilde{H}_{12} + 
2 \tilde{T}_{13} \tilde{H}_{13} + 2 \tilde{T}_{23} \tilde{H}_{23} 
\end{equation}
accounts for cross-correlation terms.

The previous expression (\ref{doro_inter_extended})  generalizes
Doroshkevich's formula (\ref{doro_standard}) to include 
the fact that halos (voids) may correspond to
maxima (minima) of the density field. 
Note that, if $r=0$, Doroshkevich's formula 
(\ref{doro_standard}) is indeed recovered -- as expected in the limit of
no correlations between the potential and density fields.
Equation (\ref{doro_inter_extended}) is written in a 
Doroshkevich-like format,  
and it is one of the key results of this paper.   

It is also possible to express (\ref{doro_inter_extended}) in terms of
the constrained eigenvalues $\zeta'$s of the ($\tilde{T}|r,\tilde{H}$) matrix.
The result is:
\begin{eqnarray}
\label{doro_eigen}
p(\tilde{\zeta}_1,
  \tilde{\zeta}_2, \tilde{\zeta}_3|r) &\equiv& p(\tilde{\lambda}_1,
  \tilde{\lambda}_2, \tilde{\lambda}_3|r, \tilde{\xi}_1,
  \tilde{\xi}_2, \tilde{\xi}_3) \\
&=& {15^3 \over 8 \sqrt {5} \pi} {1 \over (1-r^2)^3} 
 {\rm exp} \Big [-{3 \over 2 (1-r^2)} (2 K_1^2 - 5 K_2 ) \Big ] \cdot
|(\tilde{\zeta}_1-\tilde{\zeta}_2) (\tilde{\zeta}_1-\tilde{\zeta}_3) (\tilde{\zeta}_2-\tilde{\zeta}_3)| \nonumber      
\end{eqnarray}
where in terms of constrained eigenvalues:
\begin{eqnarray}
\label{eigen_doro_ext}
K_1 &=& \tilde{\zeta}_1 + \tilde{\zeta}_2 +\tilde{\zeta}_3 =   k_1 - r h_1 \\
K_2 &=&  \tilde{\zeta}_1 \tilde{\zeta}_2 + \tilde{\zeta}_1 \tilde{\zeta}_3 + \tilde{\zeta}_2 \tilde{\zeta}_3 
 = k_2 + r^2 h_2 - r h_1 k_1 + r \eta \\
\eta &=& \tilde{\lambda}_1 \tilde{\xi}_1 +  \tilde{\lambda}_2 \tilde{\xi}_2 +  \tilde{\lambda}_3 \tilde{\xi}_3 \\  
k_1 &=&  \tilde{\lambda}_1 + \tilde{\lambda}_2 + \tilde{\lambda}_3 \\
k_2 &=&  \tilde{\lambda}_1 \tilde{\lambda}_2 +  \tilde{\lambda}_1 \tilde{\lambda}_3 +  \tilde{\lambda}_2 \tilde{\lambda}_3 \\
h_1 &=&  \tilde{\xi}_1 + \tilde{\xi}_2 + \tilde{\xi}_3 \\
h_2 &=&  \tilde{\xi}_1 \tilde{\xi}_2 +\tilde{\xi}_1 \tilde{\xi}_3 + \tilde{\xi}_2 \tilde{\xi}_3
\end{eqnarray}
with the partial distributions expressed by Doroshkevich's
unconditional formulae as
\begin{equation}
p(\tilde{\lambda}_1,\tilde{\lambda}_2,\tilde{\lambda}_3) = {15^3 \over
    8 \sqrt{5} \pi} e^{\rm - {3
    \over 2} (2 k_1^2 - 5 k_2)} (\tilde{\lambda}_1-\tilde{\lambda}_2)
    (\tilde{\lambda}_1-\tilde{\lambda}_3) (\tilde{\lambda}_2-\tilde{\lambda}_3),
\label{doro_original}
\end{equation}
\begin{equation}
p(\tilde{\xi}_1,\tilde{\xi}_2,\tilde{\xi}_3) = {15^3 \over 8 \sqrt{5}
    \pi} e^{\rm - {3 \over 2} (2 h_1^2 - 5 h_2)} (\tilde{\xi}_1-\tilde{\xi}_2)
    (\tilde{\xi}_1-\tilde{\xi}_3) (\tilde{\xi}_2-\tilde{\xi}_3).
\end{equation}
and
\begin{equation}
\tilde{\zeta}_{\rm i} \equiv  (\tilde{\lambda}_{\rm i}|r, \tilde{\xi}_{\rm
  i}) = \tilde{\lambda}_{\rm i} - r \tilde{\xi}_{\rm i}
\label{constrained_eigen}
\end{equation}
which is obvious from  (\ref{K_def}),  
with $\tilde{\lambda}_{\rm i}$'s and $\tilde{\xi}_{\rm i}$'s the
corresponding unconstrained eigenvalues of the matrices $\tilde{T}$ and
$\tilde{H}$. 
Equation (\ref{doro_eigen}) is another key result of this work. This relation 
is easily derived from  (\ref{doro_inter_extended}) because the volume element in the
six-dimensional space of symmetric 
real matrices is simply given, in terms of constrained eigenvalues, by
\begin{equation}
 \prod_{\alpha} {\rm d}(\tilde{T}_{\alpha}|r,\tilde{H}_{\alpha}) =  |(\tilde{\zeta}_1 - \tilde{\zeta}_2) (\tilde{\zeta}_2 - \tilde{\zeta}_3) (\tilde{\zeta}_1 - \tilde{\zeta}_3) | {\rm d} \tilde{\zeta}_1 
{\rm d}\tilde{ \zeta}_2 {\rm d} \tilde{\zeta}_3 {\rm d} \Omega_{\rm S^3}  
\end{equation}
where ${\rm d} \Omega_{S^3} $ is the volume element of the
three-dimensional rotation group $SO(3)$, and also of the three-sphere
(see Appendix B in Bardeen et al. 1986). 
In addition, from (\ref{K_def}) it is direct to see that one has
$\tilde{T}_{\rm ii} \equiv \tilde{\lambda}_{\rm i}$ and 
$\tilde{H}_{\rm ii} \equiv \tilde{\xi}_{\rm i}$ in the system
where both $\tilde{T}$ and $\tilde{H}$ are diagonal; therefore $K_1 =
\sum_{\rm i} (\tilde{\lambda}_{\rm i} - r \tilde{\xi}_{\rm i}) \equiv
\sum_{\rm i} \zeta_{\rm i}$, with $\zeta_{\rm i}$ the constrained eigenvalues.

The fact that the system in which $\tilde{T}$
is diagonal is also the system in which $\tilde{H}$ is diagonal
has a simple geometrical explanation: in order to obtain the conditional distribution
($\tilde{T}|r,\tilde{H}$),
a standard procedure for multivariate Gaussians (when transforming from uncorrelated to correlated
variates) is to use a Cholesky
decomposition. However, this decomposition does not sample
symmetrically with respect to the principal axes. 
An alternative 
method which takes care of the alignment with the principal axes (and
used here) is the following: decompose spectrally the covariance matrix of
($\tilde{T}|r,\tilde{H}$), ${\cal A}$, i.e.
find the eigensystem $D$  and diagonal eigenvalue matrix $\Lambda$ of
the covariance matrix such that $ {\cal A} = D \cdot \Lambda
\cdot D^{\rm T}$; then set $W = D \cdot \Lambda^{1/2} Y$, with $Y$ a
set of independent zero mean unit variance Gaussians. It is direct to
show that $W$ is indeed the correct covariance matrix, and  
this scheme takes care of the alignment with the principal axes.


\subsection{The reverse joint conditional distribution of eigenvalues} \label{bbks_connection}


Equation (\ref{what_we_want}) expresses the probability of observing a
tidal field $\tilde{T}$ in regions where the curvature $\tilde{H}$ of
the density field is positive/negative (i.e. peak or dip regions). One
may be also interested in the reverse joint conditional distribution, namely
$p(\tilde{H}| r, \tilde{T} > 0 )$. Obtaining its expression is easily
achieved from the previous algebra by
applying Bayes' theorem, so that:
\begin{equation}
p(\tilde{H}| r, \tilde{T} > 0 ) = {\int_{\tilde{T} >0} p(\tilde{H}, \tilde{T}|r)~{\rm d}
  \tilde{T} \over p(\tilde{T}>0)}=
  {\int_{\tilde{T}>0} p(\tilde{T})~{\rm d} \tilde{T} \cdot
  p(\tilde{H}|r, \tilde{T}) \over \int_{\tilde{T}>0} p(\tilde{T})~{\rm d}\tilde{T}}
\label{what_we_want_bis}
\end{equation}
where now (symmetrically) 
\begin{equation}
p(\tilde{H}|r, \tilde{T}) = { p(\tilde{H}, \tilde{T}|r) \over
  p(\tilde{T})} = { p(\tilde{H}, \tilde{T}|r) \over  \int_{\rm \tilde{H}} p(\tilde{H}, \tilde{T}|r)~{\rm d} \tilde{H}} = 
{ e^{ - \rm {1 \over 2} (\tilde{H}-b')^{\rm T} \cdot {\mathscr{A}}^{-1} \cdot (\tilde{H}-b') }  \over (2 \pi)^3   \sqrt{|\mathscr{A}| } }.
\end{equation}
The distribution $p(\tilde{H}|r, \tilde{T})$ is also a multidimensional Gaussian with mean $b'$ and covariance matrix $\mathscr{A}$, where 
\begin{equation}
b' = r \tilde{T}, ~ {\mathscr{A}} = {1 \over 15} (1 -r^2) A.
\end{equation}
Along the lines of the previous calculation, it is direct to obtain
\begin{equation}
p (\tilde{H}|r, \tilde{T}) = {15^3 \over 16 \sqrt {5} \pi^3} {1 \over (1-r^2)^3} {\rm exp} \Big [-{3 \over 2 (1-r^2)} (2 H_1^2 - 5 H_2 ) \Big ]
\label{bbks_inter_extended}
\end{equation}
where in analogy with (\ref{doro_inter_extended}) one has now:
\begin{eqnarray}
\label{H_def}
H_1 &=& (\tilde{H}_{11} -r \tilde{T}_{11}) + (\tilde{H}_{22} -r \tilde{T}_{22}) + (\tilde{H}_{33} -r \tilde{T}_{33}) = h_1 - r k_1 \nonumber \\
H_2 &=& (\tilde{H}_{11} -r \tilde{T}_{11})(\tilde{H}_{22} - r \tilde{T}_{22})  + (\tilde{H}_{11} -r \tilde{T}_{11})(\tilde{H}_{33} - r \tilde{T}_{33}) \nonumber \\
&+& (\tilde{H}_{22} -r \tilde{T}_{22})(\tilde{H}_{33} - r \tilde{T}_{33}) -(\tilde{H}_{12} -r \tilde{T}_{12})^2 
-(\tilde{H}_{13} -r \tilde{T}_{13})^2 -(\tilde{H}_{23} -r \tilde{T}_{23})^2  = h_2 + r^2 k_2 - r k_1 h_1 + r \eta.
\end{eqnarray}
In addition, if $\epsilon_1, \epsilon_2, \epsilon_3$
are the constrained eigenvalues of ($\tilde{H}|r, \tilde{T}$), then 
\begin{eqnarray}
\label{bbks_eigen}
p(\tilde{\epsilon}_1,
  \tilde{\epsilon}_2, \tilde{\epsilon}_3|r) &\equiv& p(\tilde{\xi}_1,
  \tilde{\xi}_2, \tilde{\xi}_3|r, \tilde{\lambda}_1,
  \tilde{\lambda}_2, \tilde{\lambda}_3) \\
&=& {15^3 \over 8 \sqrt {5} \pi} {1 \over (1-r^2)^3} 
 {\rm exp} \Big [-{3 \over 2 (1-r^2)} (2 H_1^2 - 5 H_2 ) \Big ] \cdot
|(\tilde{\epsilon}_1-\tilde{\epsilon}_2) (\tilde{\epsilon}_1-\tilde{\epsilon}_3) (\tilde{\epsilon}_2-\tilde{\epsilon}_3)| \nonumber      
\end{eqnarray}
where in terms of constrained eigenvalues
\begin{eqnarray}
H_1 &=& \tilde{\epsilon}_1 + \tilde{\epsilon}_2 +\tilde{\epsilon}_3 =   h_1 - r k_1 \\
H_2 &=&  \tilde{\epsilon}_1 \tilde{\epsilon}_2 + \tilde{\epsilon}_1 \tilde{\epsilon}_3 + \tilde{\epsilon}_2 \tilde{\epsilon}_3 
 = h_2 + r^2 k_2 - r h_1 k_1 + r \eta, 
\end{eqnarray}
with the $\epsilon_{\rm i}$'s simply given by
\begin{equation}
\tilde{\epsilon}_{\rm i} \equiv  (\tilde{\xi}_{\rm i}|r, \tilde{\lambda}_{\rm
  i}) = \tilde{\xi}_{\rm i} - r \tilde{\lambda}_{\rm i}.
\label{constrained_eigen_bis}
\end{equation}
Note an interesting symmetry: $K_1$, $H_1$, $k_1$, $h_1$ are always the traces of
their corresponding matrices. 

New analytic formulae derived from (\ref{bbks_inter_extended}) and (\ref{bbks_eigen}), 
which generalize the work of Bardeen et
al. (1986), will be presented in a forthcoming publication.
The previous relations are
also useful in making the connection with the work of 
van de Weygaert \& Bertschinger (1996). For example, Equation (\ref{constrained_eigen_bis})
confirms their finding of the correlation between the shifted
mean values for density and shear (see for example
Equation
108 in van de Weygaert \& Bertschinger 1996, which provides the reverse mean
value of the shear given a density field shape, and their Section 4.4
for more details). 


\section{Individual conditional distributions and probabilities} \label{individual_eigenvalues}

\begin{figure}
\centering
\includegraphics[angle=0,width=1.00\textwidth]{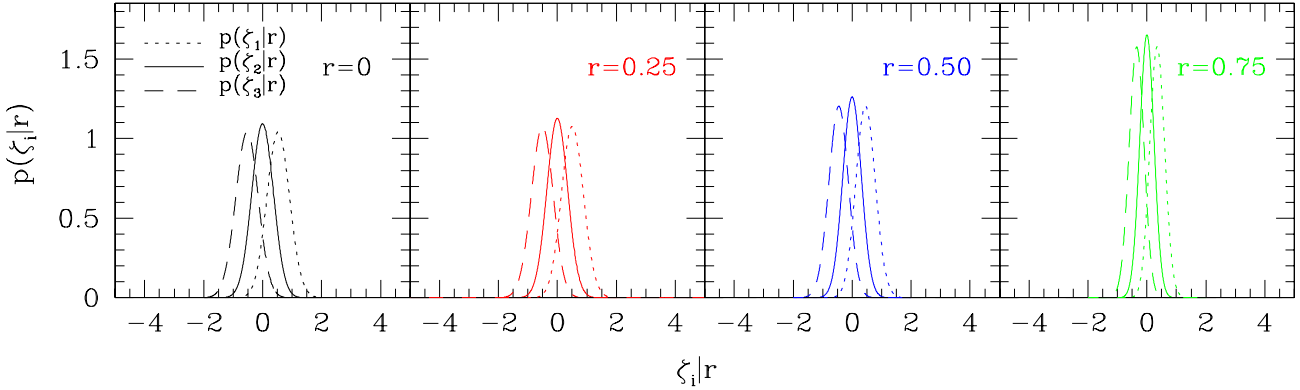}
\caption{Individual conditional distributions $p(\zeta_{\rm i}|r)$ of
  the initial
  shear field in the peak/dip picture (Equations \ref{zeta1}, \ref{zeta2} and
  \ref{zeta3}), for different values of the reduced correlation parameter
  $r$ -- as indicated in the panels.}
\label{extended_zetas}
\end{figure}

A direct application of the main formulae derived in
Section \ref{joint_eigen_new} is to compute the individual
distributions of eigenvalues (along with some other related conditional
probabilities), given the peak/dip constraint.
Starting from Equation (\ref{doro_eigen}),
with some extra complications due to the inclusion of correlation
terms, it is possible to extend the work of Lee \& Shandarin (1998)
to account for those regions where the source of the displacement is
positive (negative) -- see the Appendix \ref{Lee_Shandarin} for a
compact summary of their results.    
Introducing the following notation:
\begin{eqnarray}
l &=& {1125 \over 64 \sqrt{5} \pi} \\ 
l' &=& l \cdot (1 -r^2)^{-5/2}  \nonumber \\ 
n &=& \sqrt{3 \pi} / 12 \nonumber \\
L'(x,y|r) &=& \sqrt{1-r^2}~(x-y) (3x-y) \cdot {\rm exp} \Big [-{3
    \over (1-r^2) }\Big(x^2 -xy +{3 \over
    2}y^2 \Big ) \Big ]
\nonumber \\
N'(x,y|r) &=& (x-y)[8(1-r^2)+3(3x-y)(3y-x)]\cdot  {\rm exp}\Big[-{15
    \over 16 (1-r^2)}
  \Big (3 x^2 -2 xy + 3 y^2 \Big ) \Big ] \cdot {\rm erfc}
\Big [{\sqrt{3}
    \over 4} {1 \over \sqrt{1-r^2}} (x-3y)\Big ] \nonumber
\end{eqnarray}
and dropping the understood $\sim$ symbol (but note that the eigenvalues are still ``reduced variables''),
the two-point probability distributions of the constrained eigenvalues of
the shear field are given by:
\begin{eqnarray}
p(\zeta_1,\zeta_2|r) &=& l' \Big[ L'(\zeta_1,\zeta_2|r) + n \cdot
  N'(\zeta_1, \zeta_2|r) \Big ] \\
p(\zeta_2,\zeta_3|r) &=&l' \Big[ L'(\zeta_3,\zeta_2|r) + n \cdot
  N'(\zeta_3, \zeta_2|r) \Big ] \nonumber \\
p(\zeta_1,\zeta_3|r) &=& l' \Big \{ L'(\zeta_1,\zeta_3|r) +
  L'(\zeta_3,\zeta_1|r) + n \cdot \Big [
  N'(\zeta_1, \zeta_3|r) +  N'(\zeta_3, \zeta_1|r)  \Big ] \Big \}. \nonumber
\end{eqnarray}
Performing the various partial integrations, it is straightforward to obtain:
\begin{eqnarray}
\label{zeta1}
p(\zeta_1|r) &=& l' \Big \{ \int_{-\infty}^{\zeta_1} L'(\zeta_1,\zeta_2|r)
{\rm d} \zeta_2 + n \int_{-\infty}^{\zeta_1}
N'(\zeta_1,\zeta_2|r) {\rm d} \zeta_2 \Big \} \\
 &=& {\sqrt{5} \over 12 \pi} \Big [ {20 \over (1-r^2)}  \zeta_1 ~{\rm exp} \Big (-
  {9 \zeta_1^2 \over 2 (1-r^2)} 
  \Big ) - {\sqrt{2 \pi} \over (1-r^2)^{3/2}}~{\rm exp} \Big(- {5 \zeta_1^2 
    \over 2 (1-r^2)}  \Big
  ) \cdot {\rm erfc} \Big (-{\sqrt{2}   \zeta_1 \over \sqrt{1-r^2}}
  \Big ) [(1-r^2)-20 \zeta_1^2] \nonumber \\
&&+ {3 \sqrt{3 \pi} \over \sqrt{1-r^2}}~
       {\rm exp} \Big (- {15  \zeta_1^2 \over 4 (1-r^2)}  \Big ) {\rm erfc} \Big (
       -{\sqrt{3} \zeta_1 \over 2 \sqrt{1-r^2}} 
       \Big )        
             \Big ] \nonumber
 \end{eqnarray}

\begin{eqnarray}
\label{zeta2}
p(\zeta_2|r) &=& l' \Big \{ \int_{\zeta_2}^{+ \infty} L'(\zeta_1,\zeta_2|r)
{\rm d} \zeta_1 + n \int_{\zeta_2}^{+ \infty}
N'(\zeta_1,\zeta_2|r) {\rm d} \zeta_1 \Big \} \\
 &=& {\sqrt{15} \over 2 \sqrt{\pi}} {1 \over \sqrt{1-r^2}}
{\rm exp} \Big [- {15 \over 4}
  {\zeta_2^2 \over (1-r^2) }  \Big ] \nonumber
\end{eqnarray}

\begin{eqnarray}
\label{zeta3}
p(\zeta_3|r) &=& l' \Big \{ \int_{\zeta_3}^{\zeta_1} L'(\zeta_3,\zeta_2|r)
{\rm d} \zeta_2 + n \int_{\zeta_3}^{\zeta_1}
N'(\zeta_3,\zeta_2|r) {\rm d} \zeta_2 \Big \} \\
 &=& -{\sqrt{5} \over 12 \pi} \Big [ {20 \over (1-r^2)}  \zeta_3 ~{\rm exp} \Big (-
  {9 \zeta_3^2 \over 2 (1-r^2)} 
  \Big ) + {\sqrt{2 \pi} \over (1-r^2)^{3/2}}~{\rm exp} \Big(- {5 \zeta_3^2 
    \over 2 (1-r^2)}  \Big
  ) \cdot {\rm erfc} \Big ({\sqrt{2}   \zeta_3 \over \sqrt{1-r^2}}
  \Big ) [(1-r^2)-20 \zeta_3^2] \nonumber \\
&&- {3 \sqrt{3 \pi} \over \sqrt{1-r^2}}~
       {\rm exp} \Big (- {15  \zeta_3^2 \over 4 (1-r^2)}  \Big ) {\rm erfc} \Big (
       {\sqrt{3} \zeta_3 \over 2 \sqrt{1-r^2}} 
       \Big )        
             \Big ]. \nonumber
 \end{eqnarray}

\begin{figure}
\centering
\includegraphics[angle=0,width=1.0\textwidth]{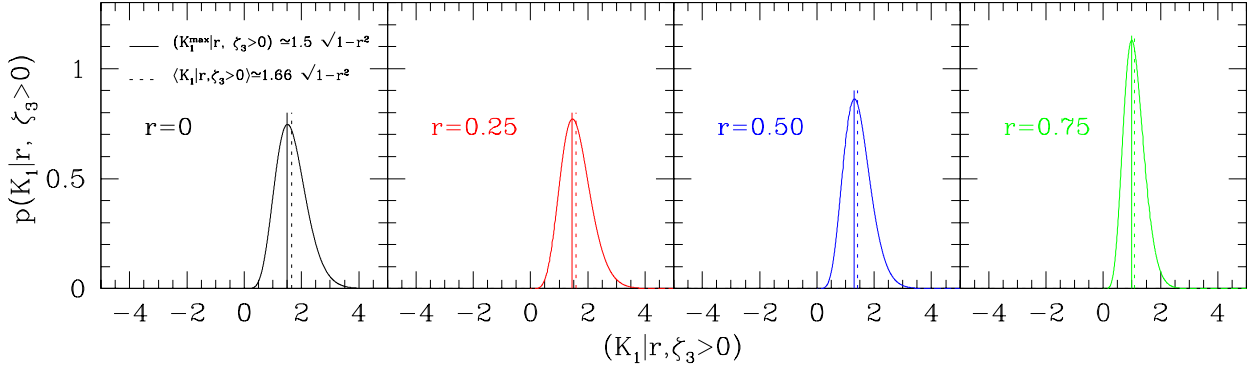}
\caption{Conditional probability distribution
$p(K_1|r, \zeta_3 >0) \equiv p(\Delta|r, \zeta_3 >0)$ given by Equation (\ref{K1_positive_eq}), for
different values of $r$, as indicated in the panels. Solid vertical
lines show the maximum of each distribution ($K_1^{\rm max} \simeq 1.5~\sqrt{1-r^2}$),
dotted lines represent their corresponding average values as given by
Equation (\ref{K1_positive_ave}).}
\label{K1_positive}
\end{figure}

\begin{figure}
\centering
\includegraphics[angle=0,width=0.5\textwidth]{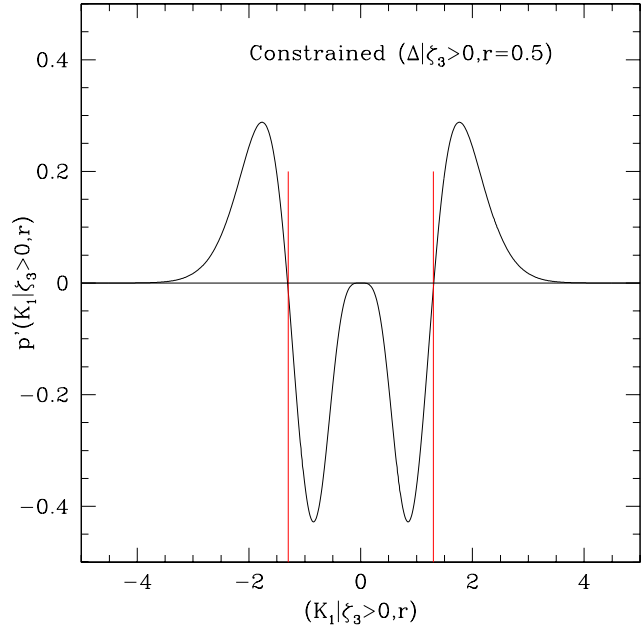}
\caption{Determination of the maxima of (\ref{K1_positive_eq}). The
  derivative $p'(K_1|\zeta_3>0,r)$ is shown, for an arbitrary value of
  the reduced correlation $r$, i.e. $r=0.5$. The maxima correspond to those
  points where $K_1^{\rm
  max} \simeq 1.5~\sqrt{1-r^2}$, indicated with solid vertical lines
  in the figure.}
\label{K1_positive_max}
\end{figure}

Plots of these distributions are shown in Figure \ref{extended_zetas},
for different values of the reduced correlation $r$.
Note the symmetry between $p(\zeta_1|r)$ and
$p(\zeta_3|r)$, evident from (\ref{zeta1}) and (\ref{zeta3}). 
In particular, $p(\zeta_2|r)$ is simply a Gaussian with
variance $2 (1-r^2)/15$, and so the conditional distribution $p(\lambda_2|r,\xi_2)$ is a
Gaussian with the same variance and shifted mean $r \xi_2$ (recall
that $\zeta_2 = \lambda_2 - r \xi_2$).
In addition, it is easy to show that, with $\Delta = \zeta_1 +
\zeta_2 + \zeta_3 \equiv K_1$, one obtains
\begin{eqnarray}
p(\Delta, \zeta_3|r) &=& {3 \sqrt{5} \over 16 \pi} {1 \over (1-r^2)^2} \Big [15 \Delta^2 -
  90 \Delta \zeta_3 + 135 \zeta_3^2 -8 (1-r^2)\Big ] \cdot {\rm exp} \Big
  [- {3 \over 8 (1-r^2)} (3 \Delta^2 -10 \Delta \zeta_3 + 15 \zeta_3^2) \Big]   \nonumber\\
&& + {3 \sqrt{5} \over 2 \pi} {1 \over (1-r^2)} {\rm exp} \Big [-{3
  \over (1-r^2)} \Big ( \Delta^2 -5
  \Delta \zeta_3 +{15 \over 2} \zeta_3^2 \Big )  \Big ].
\end{eqnarray}
By integrating the previous equation over $\zeta_3$, it is direct
to verify that $p(\Delta|r) \equiv p(K_1|r)$ is a Gaussian with
zero mean and variance $(1-r^2)$, namely:
\begin{equation}
p(\Delta|r) = {1 \over \sqrt{2\pi (1-r^2)}} \cdot {\rm
  exp} \Big [- {\Delta^2 \over 2 (1-r^2)} \Big ] \equiv p(K_1|r).
\end{equation}  
Since $K_1 = k_1-r h_1$ (i.e. Eq. \ref{eigen_doro_ext}),
the previous expression implies that $p(k_1|r, h_1)$ is therefore a Gaussian with mean $rh_1$ and
variance $(1-r^2)$. 
In addition:
\begin{eqnarray}
p(\zeta_1>0|r) &=& {23 \over 25},~~~\langle \zeta_1|r \rangle = {3
  \over \sqrt{10 \pi}}\sqrt{1-r^2},~~~~ \sigma^2_{\rm \zeta_1|r} = {13 \pi -27
  \over 30 \pi} (1-r^2)\\
p(\zeta_2>0|r) &=& {1 \over 2},~~~~~\langle \zeta_2|r \rangle = 0,~~~~~~~~~~
\sigma^2_{\rm \zeta_2|r} = {2 \over 15} (1-r^2)\\ 
p(\zeta_3>0|r) &=& {2 \over 25},~~~\langle \zeta_3|r \rangle = -{3
  \over \sqrt{10 \pi}}\sqrt{1-r^2},~~ \sigma^2_{\rm \zeta_3|r} = {13 \pi -27
  \over 30 \pi}(1-r^2).
\end{eqnarray}
Moreover, the probability distribution of $\Delta$ confined in the regions with
$\zeta_3 > 0$ is:
\begin{eqnarray}
\label{K1_positive_eq}
p(\Delta |r, \zeta_3 >0) &=& {p(\Delta, \zeta_3>0|r) \over
  p(\zeta_3>0)} = {\int_{0}^{\rm \Delta/3}  p(\Delta, \zeta_3|r)
  {\rm d} \zeta_3 \over
  p(\zeta_3>0)} \\
&=& -{75 \sqrt{5} \over 8 \pi } {\Delta \over (1-r^2)}~{\rm exp} \Big( - {9 \over 8}
  {\Delta^2 \over (1-r^2)} \Big) + \nonumber \\
&+& {25 \over 4 \sqrt{2 \pi} } {1 \over
  \sqrt{1-r^2}}~{\rm exp} \Big ( -
  {\Delta^2 \over 2(1-r^2)} \Big ) \Big[{\rm erf}\Big ({\sqrt{10} \Delta
  \over 4 \sqrt{1-r^2}} \Big )+ {\rm erf}\Big ({\sqrt{10} \Delta
  \over 2 \sqrt{1-r^2}} \Big )    \Big]. \nonumber
\end{eqnarray}
Using the previous formula, it is direct to prove that
\begin{equation}
\langle \Delta |r, \zeta_3 > 0 \rangle = {25 \sqrt{10} \over 144
  \sqrt{\pi}} (3 \sqrt{6} -2)~ \sqrt{1-r^2}\simeq 1.6566~\sqrt{1-r^2}.
\label{K1_positive_ave}
\end{equation}
Figure \ref{K1_positive} shows  the conditional probability distribution
$p(K_1|r, \zeta_3 >0) \equiv p(\Delta|r, \zeta_3 >0)$ (i.e. Equation \ref{K1_positive_eq}), for
different values of $r$. The dotted vertical lines display the average
values of those functions, as given in (\ref{K1_positive_ave}), while the
vertical solid lines indicate their maximum values. 
In fact, it is easy to see that the maximum of $p(\Delta|r,\zeta_3>0)$
is reached when $\Delta \simeq 1.5~\sqrt{1-r^2}$. This result is readily obtained by
computing the derivative of $p(\Delta|r,\zeta_3>0)$, and by finding the corresponding zeroes.
An example of this calculation is shown in Figure
\ref{K1_positive_max}, where an arbitrary value of $r=0.5$ is considered. 

Finally, the probability that a given region with density $\Delta$
will have all positive eigenvalues $\zeta$'s is:
\begin{equation}
p(\zeta_3 > 0| \Delta, r) = {p(\Delta, \zeta_3>0|r) \over
  p(\Delta|r)} 
= - {3 \sqrt{10} \over 4 \sqrt{\pi} } {\Delta \over \sqrt{1-r^2}} ~{\rm exp} \Big (- {5
  \over 8} {\Delta^2 \over (1-r^2)}  \Big ) + {1 \over 2} \Big [ {\rm erf}
  \Big({\sqrt{10} \Delta \over 4 \sqrt{1-r^2}}
  \Big )  + {\rm erf}
  \Big({\sqrt{10} \Delta \over 2 \sqrt{1-r^2}}
  \Big ) \Big ].  \nonumber
\end{equation}

Note that, in absence of correlations between the potential and
density fields (i.e. when $r=0$), all the previous expressions reduce
consistently to the unconditional limit of Lee \& Shandarin (1998)
-- see again Appendix \ref{Lee_Shandarin}.
Several other results on probability distributions along these lines were also already derived 
by van de Weygaert \& Bertschinger (1996): here we confirm their
findings that the inclusion of the condition for a peak or dip in the
Gaussian density field involves a $1/\sqrt{1-r^2}$ factor, and that
the conditional distributions have shifted mean and reduced variance
(see again their Section 4).



\section{CONCLUSION} \label{conclusion}

Since the initial shear field associated with Gaussian statistics plays a major role in the formation of
large scale structures, considerable analytic work has been based 
on the joint distribution of its eigenvalues -- i.e. 
Doroshkevich's formulae (Equation \ref{doro_standard}).
However, Doroshkevich's equations neglect the fact that halos (voids) may correspond to maxima (minima) of the density field. 
The main goal of this work was to provide new analytic expressions, in the context of the peak/dip
picture (Bardeen et al. 1986; Bond et al. 1991),
which include the fact that the eigenvalues of the linear shear field
are related to regions where the source of the displacement is positive (negative).    
These new conditional probabilities, derived in Section \ref{new_formula}, are Equations (\ref{doro_inter_extended}), (\ref{doro_eigen}),
(\ref{bbks_inter_extended}) and (\ref{bbks_eigen}): they represent the main results
of this paper. Written in Doroshkevich-like format,
they naturally reduce to Doroshkevich's (1970) unconditional
relations in the limit of no correlation between the potential and
the density fields (i.e. when $r=0$). 
As a first direct application of (\ref{doro_eigen}), the 
individual conditional distributions of eigenvalues were obtained
in Section \ref{individual_eigenvalues}; these relations extend some previous work by Lee 
\& Shandarin (1998) -- see Appendix \ref{Lee_Shandarin} for a compact
summary of their results.
Much more analytic work can be carried out using these new formulae,
especially in connection with the statistics of peaks developed by
Bardeen et al. (1986): their calculations can be extended within
this framework (see Section
\ref{bbks_connection}),  and results of this extension will be presented in
a forthcoming publication -- along with more insights on the equations derived in Section \ref{joint_eigen_new}.

To obtain the conditional distribution $p(\tilde{T}|r,\tilde{H}>0)$ (i.e. Equation \ref{what_we_want}), in principle
one needs to sample numerically the probability (\ref{cond}) as done in Lavaux \& Wandelt (2010). However,
the new analytic results of this paper suggest a simpler generalized excursion set algorithm, 
which will be also presented in a forthcoming study. The algorithm allows for a fast 
sampling of (\ref{cond}), and permits to test the new relations
derived here
(i.e. Sections \ref{new_formula} and \ref{individual_eigenvalues}, and Appendix \ref{Lee_Shandarin})
against mock data.
In addition, along these lines it is possible to extend the main shape distributions
involved in the triaxial formation of nonlinear structures
(i.e. ellipticity $e$, prolateness $p$, axis ratios $\nu$ and $\mu$, etc.).
Halos and voids are in fact triaxial rather than spherical 
(i.e. Rossi, Sheth \& Tormen 2011), and the 
initial shear field plays a crucial role in their formation. For example, in the ellipsoidal
collapse framework the virialization condition depends on 
ellipticity and prolateness, which are directly related to the eigenvalues of
the external tidal field. Starting from
$p(\zeta_1,\zeta_2,\zeta_3|r)$, it is then straightforward to
turn this joint conditional distribution into $p(e,p,\delta|r)$ -- to include the peak
constrain, -- and then to characterize $p(e|r)$, $p(p|r)$, $p(\mu|r)$, $p(\nu|r)$ and so forth. 
Applications to halos and voids will be also presented next,
including implications for the skeleton of the cosmic web.
The extension to non-Gaussian fields, along the lines of Lam et
al. (2009), is a natural follow-up of this work and will be presented in a
forthcoming publication as well. In the context of cosmic voids, other
interesting uses of the new formulae
 (\ref{doro_inter_extended}), (\ref{doro_eigen}),
(\ref{bbks_inter_extended}) and (\ref{bbks_eigen}) involve the
Monge-Amp\`{e}re-Kantorovitch reconstruction procedure -- see for example Lavaux \& Wandelt (2010). 

The analytic framework described here can be also useful in 
several observationally-oriented applications, and in
particular for developing 
algorithms to find and classify structures in the cosmic web. 
For example, Bond, Strauss \& Cen (2010) presented an algorithm that
uses the eigenvectors of the Hessian matrix of the smoothed galaxy
distribution to identify individual filamentary structures.
They used the distribution of the Hessian eigenvalues of the smoothed
density field on a grid to study clumps, filaments and walls.
Other possibilities include a 
web classification based on the multiscale analysis of the Hessian
matrix of the density field (Aragon-Calvo et al. 2007), the skeleton
analysis (Novikov et al. 2006), as well as
a morphological (Zeldovich-based) classification (for instance Klypin \& Shandarin 1983; Forero-Romero et al. 2009).
More generally, the fact that the eigenvalues of the Hessian matrix can be used to
discriminate different types of structure in a particle distribution
is fundamental to a number of structure-finding algorithms
(Forero-Romero et al. 2009), shape-finders algorithms (Sahni et
al. 1998), and structure reconstruction on the basis of tessellations (Schaap \&
van de Weygaert 2000; Romano-Diaz \& van de Weygaert 2007), etc.
In addition, 
the classification of different environments should provide a framework for
studying the environmental dependence of galaxy formation (see for
example Blanton et al. 2005).
To this end, recently Park,
Kim and Park (2010) extended the concept of galaxy environment to the gravitational potential
and its functions -- as the shear tensor. They studied how to 
accurately estimate the gravitational potential from an observational sample
finite in volume, biased due to galaxy biasing, and subject to
redshift space distortions, by inspecting the dependence of dark
matter halo properties on environmental parameters (i.e. local
density, gravitational potential, ellipticity and prolateness of the
shear tensor). It would be interesting to interpret their results
with the theoretical
formalism developed here, and ultimately to
study the gravitational potential directly from a real dataset such as
the SDSS Main Galaxy sample within this framework.

The formalism presented in this paper is restricted to one scale
(i.e. peaks and dips in the density field, as in Bardeen et al. 1986), but
the extension to a multiscale \textit{peak-patch} approach along the lines of Bond \& Myers
(1996) is duable and subject of ongoing work. This will 
allow to account for the role of the peculiar
 gravity field itself, an important aspect not considered here but discussed for example in 
 van de Weygaert \& Bertschinger (1996). 
In fact, these authors introduced the peak constraints
 to describe the density field in the immediate surroundings of a
 peak, and then addressed the constraints on the gravitational
 potential perturbations; in particular, they constrained the peculiar
 gravitational acceleration at the position of the peak itself, in
 addition to characterizing the tidal field around the peak. 
 Including all these effects in our formalism
 is ongoing effort.
Finally, the
interesting and more complex question of the
local expected density field alignment/orientation distribution as a
function of the local field value (or the other way around -- see Bond
1987; Lee \& Pen 2002; Porciani et al. 2002;  Lee 2011) can be addressed within this framework, and is left to future studies. 



\section{ACKNOWLEDGMENTS}

I would like to thank Ravi K. Sheth especially for bringing to my attention the
work of Lee \& Shandarin (1998), and Changbom Park for his always
wise advices -- along with numerous discussions.
I would also like to thank the referee, Rien van de Weygaert, for his
insightful report and useful suggestions -- which have been all included in the paper. 
Part of this work was carried out in June 2011 during the
``APCTP-IEU Focus Program on Cosmology and Fundamental Physics'' at
Postech in Pohang (Korea), and completed during the ``Cosmic Web Morphology and  Topology''
meeting  at the Nicolaus Copernicus Astronomical Center in Warsaw (Poland), on July 12-17, 2011 -- where I had stimulating conversations
with Bernard Jones and Sergei Shandarin. I would like to thank all the organizers of these workshops, and in
particular Changrim Ahn for the former and Wojciech A. Hellwing, Rien van de Weygaert
and Changbom Park for the latter. 





\appendix


\section{Individual unconditional distributions and probabilities} \label{Lee_Shandarin}

Starting from Doroshkevich's formulae
for the unconditional distribution of the ordered eigenvalues $\lambda_1 \ge \lambda_2 \ge \lambda_3$ (Eq. \ref{doro_original}),  Lee \& Shandarin (1998)
derived a set of useful probability functions.
Their expressions can be considerably
simplified by using various symmetries, and
by introducing the following notation:
\begin{eqnarray}
l &=& {1125 \over 64 \sqrt{5} \pi} \\ 
n &=& \sqrt{3 \pi} / 12 \nonumber \\
L(x,y) &=& (x-y) (3x-y) \cdot {\rm exp} \Big [-3\Big(x^2 -xy +{3 \over
    2}y^2 \Big ) \Big ]
\nonumber \\
N(x,y) &=& (x-y)[8+3(3x-y)(3y-x)]\cdot {\rm exp}\Big[-{15 \over 16}
  \Big (3 x^2 -2 xy + 3 y^2 \Big ) \Big ] \cdot {\rm erfc}
\Big [{\sqrt{3}
    \over 4} (x-3y)\Big ]. \nonumber
\end{eqnarray}
Dropping the understood $\sim$ symbols for the eigenvalues (although all the quantities are
still ``reduced'', i.e. the $\sigma$ dependence is not shown),
the two-point probability distributions are:
\begin{eqnarray}
\label{ls1998}
p(\lambda_1,\lambda_2) &=& l \Big[ L(\lambda_1,\lambda_2) + n \cdot
  N(\lambda_1, \lambda_2) \Big ] \\
p(\lambda_2,\lambda_3) &=&l \Big[ L(\lambda_3,\lambda_2) + n \cdot
  N(\lambda_3, \lambda_2) \Big ] \\
p(\lambda_1,\lambda_3) &=& l \Big \{ L(\lambda_1,\lambda_3) +
  L(\lambda_3,\lambda_1) + n \cdot \Big [
  N(\lambda_1, \lambda_3) +  N(\lambda_3, \lambda_1)  \Big ] \Big \}.
\end{eqnarray}
Performing the various integrations:
\begin{eqnarray}
p(\lambda_1) &=& l \Big \{ \int_{-\infty}^{\lambda_1} L(\lambda_1,\lambda_2)
{\rm d} \lambda_2 + n \int_{-\infty}^{\lambda_1}
N(\lambda_1,\lambda_2) {\rm d} \lambda_2 \Big \} \\
 &=& {\sqrt{5} \over 12 \pi} \Big [ 20 \lambda_1 ~{\rm exp} \Big (-
  {9 \over 2} \lambda_1^2
  \Big ) - \sqrt{2 \pi}~{\rm exp} \Big(- {5 \over 2} \lambda_1^2  \Big
  ) {\rm erfc} (-\sqrt{2} \lambda_1) (1-20 \lambda_1^2) + 3 \sqrt{3 \pi}~
       {\rm exp} \Big (- {15 \over 4} \lambda_1^2 \Big ) {\rm erfc} \Big (
       -{\sqrt{3} \over 2} \lambda_1
       \Big )        
             \Big ] \nonumber
 \end{eqnarray}

\begin{eqnarray}
p(\lambda_2) &=& l \Big \{ \int_{\lambda_2}^{+ \infty} L(\lambda_1,\lambda_2)
{\rm d} \lambda_1 + n \int_{\lambda_2}^{+ \infty}
N(\lambda_1,\lambda_2) {\rm d} \lambda_1 \Big \} \\
 &=& {\sqrt{15} \over 2 \sqrt{\pi}} {\rm exp} \Big [- {15 \over 4}
  \lambda_2^2  \Big ] \nonumber
\end{eqnarray}

\begin{eqnarray}
p(\lambda_3) &=& l \Big \{ \int_{\lambda_3}^{\lambda_1} L(\lambda_3,\lambda_2)
{\rm d} \lambda_2 + n \int_{\lambda_3}^{\lambda_1}
N(\lambda_3,\lambda_2) {\rm d} \lambda_2 \Big \} \\
 &=& - {\sqrt{5} \over 12 \pi} \Big [ 20 \lambda_3 ~{\rm exp} \Big (-
  {9 \over 2} \lambda_3^2
  \Big ) + \sqrt{2 \pi}~{\rm exp} \Big(- {5 \over 2} \lambda_3^2  \Big
  ) {\rm erfc} (\sqrt{2} \lambda_3) (1-20 \lambda_3^2) - 3 \sqrt{3 \pi}~
       {\rm exp} \Big (- {15 \over 4} \lambda_3^2 \Big ) {\rm erfc} \Big (
       {\sqrt{3} \over 2} \lambda_3
       \Big )        
             \Big ]. \nonumber
 \end{eqnarray}
Note the symmetry between $p(\lambda_1)$ and
$p(\lambda_3)$.
In addition, with $\delta = \lambda_1 +
\lambda_2 + \lambda_3 \equiv k_1$, one gets
\begin{equation}
p(\delta, \lambda_3) = {3 \sqrt{5} \over 16 \pi} \Big [15 \delta^2 -
  90 \delta \lambda_3 + 135 \lambda_3^2 -8 \Big ]\cdot {\rm exp} \Big
  [- {3 \over 8} (3 \delta^2 -10 \delta \lambda_3 + 15 \lambda_3^2) \Big] + {3 \sqrt{5} \over 2 \pi} {\rm exp} \Big [-3 \Big ( \delta^2 -5
  \delta \lambda_3 +{15 \over 2} \lambda_3^2 \Big )  \Big ].
\end{equation}
By integrating the previous expression over $\lambda_3$, it is direct
to verify that $p(\delta)$ is a zero-mean unit-variance
Gaussian distribution. In addition:  
\begin{eqnarray}
p(\lambda_1>0) &=& {23 \over 25},~~~\langle \lambda_1 \rangle = {3
  \over \sqrt{10 \pi}},~~~ \sigma^2_{\rm \lambda_1} = {13 \pi -27
  \over 30 \pi}\\
p(\lambda_2>0) &=& {1 \over 2},~~~~\langle \lambda_2 \rangle = 0,~~~~~~~~~~
\sigma^2_{\rm \lambda_2} = {2 \over 15} \\ 
p(\lambda_3>0) &=& {2 \over 25},~~~\langle \lambda_3 \rangle = -{3
  \over \sqrt{10 \pi}},~~~ \sigma^2_{\rm \lambda_3} = {13 \pi -27
  \over 30 \pi}.
\end{eqnarray}
Moreover, the probability distribution of $\delta$ confined in the regions with
$\lambda_3 > 0$ is:
\begin{eqnarray}
p(\delta | \lambda_3 >0) &=& {p(\delta, \lambda_3>0) \over
  p(\lambda_3>0)} = {\int_{0}^{\rm \delta/3}  p(\delta, \lambda_3)
  {\rm d} \lambda_3 \over
  p(\lambda_3>0)} \\
&=& -{75 \sqrt{5} \over 8 \pi } \delta~{\rm exp} \Big( - {9 \over 8}
  \delta^2 \Big) + {25 \over 4 \sqrt{2 \pi} }~{\rm exp} \Big ( -
  {\delta^2 \over 2} \Big ) \Big[{\rm erf}\Big ({\sqrt{10} \delta
  \over 4} \Big )+ {\rm erf}\Big ({\sqrt{10} \delta
  \over 2} \Big )    \Big]. \nonumber
\end{eqnarray}
Using the previous formula, it is direct to show that
\begin{equation}
\langle \delta |\lambda_3 > 0 \rangle = {25 \sqrt{10} \over 144
  \sqrt{\pi}} (3 \sqrt{6} -2) \simeq 1.6566.
\label{delta_positive_ls}
\end{equation}
It is also easy to see that the maximum of $p(\delta|\lambda_3>0)$
is reached when $\delta \simeq 1.5$: this can be readily achieved by
computing the derivative of the previous expression and by finding
the corresponding zeroes, as done in the main text (see Figure \ref{K1_positive_max}).  
Finally, the probability that a given region with density $\delta$
will have all positive eigenvalues is:
\begin{equation}
p(\lambda_3 > 0| \delta) = {p(\delta, \lambda_3>0) \over p(\delta)}
= - {3 \sqrt{10} \over 4 \sqrt{\pi} } \delta~{\rm exp} \Big (- {5
  \over 8} \delta^2  \Big ) + {1 \over 2} \Big [ {\rm erf}
  \Big({\sqrt{10} \delta \over 4}
  \Big )  + {\rm erf}
  \Big({\sqrt{10} \delta \over 2}
  \Big ) \Big ].  \nonumber
\end{equation}


\label{lastpage}


\end{document}